# So you think you can DAS?
## A viewpoint on delay-and-sum beamforming

Vincent Perrot, Maxime Polichetti, François Varray, Damien Garcia

*Abstract* – Delay-and-sum (DAS) is the most widespread digital beamformer in high-frame-rate ultrasound imaging. Its implementation is simple and compatible with real-time applications. In this viewpoint article, we describe the fundamentals of DAS beamforming. The underlying theory and numerical approach are detailed so that users can be aware of its functioning and limitations. In particular, we discuss the importance of the $f$-number and speed of sound on image quality, and propose one solution to set their values from a physical viewpoint. We suggest determining the $f$-number from the directivity of the transducer elements and the speed of sound from the phase dispersion of the delayed signals. Simplified Matlab codes are provided for the sake of clarity and openness. The effect of the $f$-number and speed of sound on the lateral resolution and contrast-to-noise ratio was investigated *in vitro* and *in vivo*. If not properly preset, these two factors had a substantial negative impact on standard metrics of image quality (namely **CNR** and **FWHM**). When beamforming with DAS *in vitro* or *in vivo*, it is recommended to optimize these parameters in order to use it wisely and prevent image degradation.

## 1. Introduction

DELAY-AND-SUM is the most basic digital beamformer for medical ultrasound imaging. Because of its simplicity and efficiency, it is very likely the most widely used in high-frame-rate (ultrafast) ultrasound. Before being used for ultrasound imaging [1], the technique of delay-and-sum (DAS) has historically been linked to ground-based and airborne radar as well as telecommunication [2]. It originates from steerable array antennas in shortwave communication [3]. Similar to a medical ultrasound transducer, an antenna array is defined as a group of connected individual antennas (or elements) that operate together. In radar and ultrasonic imaging, a *phased array* refers to a computer-controlled array that is electronically steered to transmit or receive in different directions without moving the unit elements. Although the term is generally reserved for cardiac application, most transducers used in ultrasound medical imaging are phased arrays since the delays can be modified electronically in transmission or reception.

*Multidisciplinary DAS* – In telecommunication, in contrast to ultrasound imaging, the involved travel distances are very large compared to the size of the receiver. A telecommunication antenna then receives quasi-planar waves only. An array of equally spaced unit antennas can be steered by phase-shifting to collect a radio wavefront from one direction. The radio waves of the different unit antennas are time-shifted with predefined *delays* (the greater the delays, the greater the receive angle) so that they can be *summed* to increase the signals in the desired receive direction while attenuating those from undesired directions [3]. This describes the basic concept of delay-and-sum. Computer-controlled antenna technology based on DAS has been the subject of intensive research for military radars and sonars [4], as it allowed antennas to be quickly steered to detect airplanes, submarines, and missiles. The DAS principle had also been one of the most useful techniques in seismological research. It allowed observation of small seismic events by connecting a large number of horizontally distributed seismic sensors [5]. In geophysics, DAS is often referred to as "diffraction summation". It was the first computer implementation of seismic data analysis [6]. A similar technique that incorporates amplitude correction and frequency-dependent phase correction, called "Kirchhoff migration", was then proposed and is still in common use [7]. Amplitude correction due to wave spreading is not a major concern in ultrasound imaging since the recorded signals can be time-gain compensated [8]. In addition, unlike geophysical signals, medical ultrasound signals are narrow-band, so there is little need for frequency-dependent phase correction. An important aspect of geophysics is the wide range of propagation speeds, which has promoted the development of advanced reconstruction techniques [6]. Although the study of speed heterogeneity may be valuable in some ultrasound applications [9], [10], the speed of sound is in most situations considered homogeneous in soft tissue and assumed to be equal to 1540 m/s. As we will see, this fixed value can occasionally be suboptimal.

*DAS position in ultrafast ultrasound imaging* – With the access to open-architecture ultrasound scanners and their raw data [11], ultrafast ultrasound imaging has quickly gained popularity in our community. Although the content of the manuscript could also be derived for focused beams (see Appendix 6.A), we opted for diverging [12] and plane waves to be in line with the most recent literature. "Ultrafast imaging" might be a misnomer; however, we keep this abusive terminology because

This work was supported in part by the LABEX CELYA (ANR-10-LABX-0060) and LABEX PRIMES (ANR-10-LABX-0063) of Université de Lyon, program "Investissements d'Avenir" (ANR-16-IDEX-0005).

All the authors are with CREATIS, CNRS UMR 5220, INSERM U1206, Université Lyon 1, INSA Lyon, Université Jean Monnet Saint-Etienne (e-mail: damien.garcia@inserm.fr; garcia.damien@gmail.com).

unfocused beams are generally used for high-frame-rate purposes. Even if DAS remains a widespread beamformer in high-frame-rate ultrasound imaging, there is a continuing interest in the development of alternative beamforming methods. In particular, a number of recent studies have focused on adaptive beamformers that aim to improve contrast and spatial resolution provided by DAS. A series of popular adaptive beamforming techniques are discussed in [13]. In such investigations, the DAS is most often chosen as the substandard reference method. However, as it will be explained and illustrated, care must be taken to implement it correctly. In particular, special attention must be paid to the speed of sound and the receive aperture used during beamforming. A well-implemented DAS can have a significant impact on image quality. It is highly recommended to use it properly when comparing it with other beamformers. Our objective is here to dissect the DAS by providing a detailed theoretical and pedagogical overview and suggesting some tricks for appropriate and efficient utilization.

The advantages of the DAS are many: 1) DAS is based on basic concepts of wave propagation (linearity, straight-ray propagation, weak backscattering); 2) its implementation is simple and can be parallelized; 3) it is numerically robust, fast and compatible with real-time applications; and 4) because it is data-independent, it preserves the temporal coherence and statistical properties of the real envelopes [14], [15]. The implementation of DAS in the space domain, as opposed to frequency-based approaches [16], [17], enables the notion of $f$-number and directivity, both of which will be discussed in this article. In this paper, we put DAS-based beamforming in the context of high-frame-rate imaging before detailing and exemplifying its specificities. We then show the effects of the receive aperture and speed of sound on image quality and propose potential solutions to adjust these parameters. In particular, we explain how to determine a proper receive aperture ($f$-number) from the directivity of the array elements. A method is also proposed to optimize the speed of sound. It is shown that these two fundamental aspects can significantly improve the quality of DAS-derived images. Simplified short Matlab codes are provided in the appendix for the sake of clarity and pedagogy. Complete open-source Matlab codes can be found in the MUST Matlab UltraSound Toolbox released by D. Garcia and downloadable from www.biomecardio.com/MUST.

## 2. DAS in depth

### A. Pulse-echo with a virtual point source

High-frame-rate ultrasound is most often used in conjunction with the emission of circular or plane waves. The purpose is to insonify a large region of interest with a wide wavefront so that a complete image can be obtained with a single transmission. In practice, however, a number of transmissions are necessary to allow the corresponding backscattered signals to be combined to get a high-quality ultrasound image [19]. Assuming that the insonified medium is essentially composed of pointlike Rayleigh backscatters, the latter behave as monopole sources when they are reached by the wavefront. These secondary sources re-emit the signals quasi-uniformly in all directions (spherical waves). Forming an image from sound requires estimating the round-trip traveltimes of the wavefronts towards and from all scatterers. They can be explicitly expressed when transmitting circular and plane waves, as now explained.

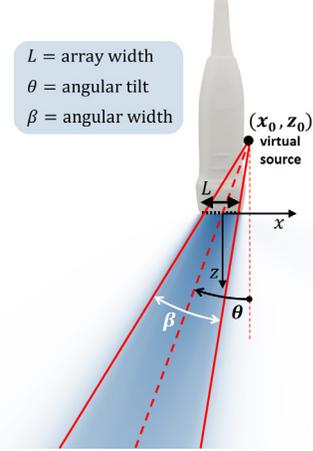

Fig. 1. Parameters that define a circular-wave transmit. It becomes a $\theta$-tilted plane-wave transmit when $\beta \to 0^+$. Figure adapted from *[20]*.

To make things simpler, we focus on rectilinear arrays (for which the positions of the elements satisfy $z = 0$) because the use of unfocused waves is still marginal with convex transducers [21]. The Cartesian coordinate system associated with a uniform linear array (ULA) is conventionally defined as follows: the $x$-axis is parallel to the transducer and points from the leftmost to the rightmost element (Fig. 1), and $x = 0$ at the center of the ULA. The $z$-axis is perpendicular to the ULA, points downward, and $z = 0$ at the level of the ULA. Note that what we refer to as a ULA is not necessarily the complete transducer array (i.e. when using a full aperture), but can be a subarray (i.e. when using a sub-aperture). Let $p$ denote the pitch of the ULA, i.e. the center-to-center distance between two adjacent elements, and $N_e$ the number of elements. The center-to-center distance from the first to the last element of the ULA therefore is $L = (N_e - 1)p$. The coordinates of the centers $(x_i, z_i)$ of the elements are given by

$$\begin{cases} x_i = \dfrac{p}{2}(2i - N_e - 1) \\ z_i = 0 \end{cases} \text{ with } i = 1 \dots N_e. \quad (1)$$

We assume that the speed of sound ($= c$) in the medium is uniform. The two-way traveltimes of the wavefronts, from the transducer to a scatterer of coordinate $\boldsymbol{X_s} = (x_s, z_s)$, and back from the scatterer to an element #$i$ of the transducer, are then defined by

$$\tau_i(\boldsymbol{X_s}) = \frac{d_{\text{TX}}(\boldsymbol{X_s}) + d_{\text{RX}}(\boldsymbol{X_s}, x_i)}{c} - t_0. \quad (2)$$

$d_{\text{TX}}$ and $d_{\text{RX}}$ are the transmit and receive distances that are described below. The parameter $t_0$ is the start time of the acquisition. It can be used to reduce the volume of the acquired data by considering only the region of interest (often useful in color Doppler, when the region of interest is a few centimeters from the probe). A factor that $t_0$ can take into account is also the different speed of sound through the acoustic lens of the transducer. The lens adds a supplementary propagation time that can be corrected by an additional $t_0$; for instance, Verasonics scanners perform this correction by default for non-custom probe if the lens correction is defined by the manufacturer. Finally, $t_0$ can correct delays due to the pulse length.

The receive distance $d_{RX}$ represents the distance traveled by the spherical wavefront generated by the point scatterer, from the scatterer to element #$i$. It is given by

$$d_{RX}(X_s, x_i) = \sqrt{(x_i - x_s)^2 + z_s^2}\ . \quad (3)$$

The transmit distance $d_{TX}$ represents the distance traveled by the wavefront from the ULA to a given point scatterer. In high-frame-rate ("ultrafast") ultrasound imaging, it is common to use circular or plane waves as they can be easily designed. As we will see below, a plane wave is a limiting case of a circular wave. We define a circular-wave transmit by its tilt $\theta \in ]\frac{-\pi}{2}, \frac{\pi}{2}[$ and its angular width $\beta \in ]0, \pi[$, both represented in Fig. 1. The angle $\beta$ is the angle between the two lines passing through the virtual source and the two centers of the edge elements. The tilt angle $\theta$ is measured counterclockwise with respect to the $z$-axis. Note that this angle $\theta$ is $> \pi$ for a focused beam. Trigonometric manipulations can provide the coordinates of the corresponding virtual point source:

$$\begin{cases} x_0 = \dfrac{L}{2}\dfrac{\sin(2\theta)}{\sin(\beta)}, \\ z_0 = -\dfrac{L}{2}\dfrac{\cos(\beta) + \cos(2\theta)}{\sin(\beta)}\ . \end{cases} \quad (4)$$

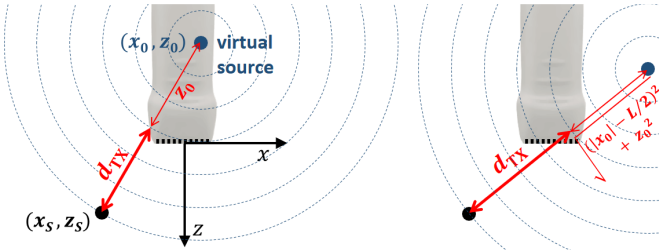

Fig. 2. Illustration of the transmit distance $d_{TX}$ for circular wave imaging, as given by Eq. (5). $L$ stands for the array aperture. $(x_s, z_s)$ are the coordinates of one scatterer.

In circular-wave imaging, the transmit distance $d_{TX}$ can be written as a function of the virtual point source as follows:

$$\begin{aligned} d_{TX}(X_s) &= \sqrt{(x_s - x_0)^2 + (z_s - z_0)^2} \\ &\quad - \sqrt{H(|x_0| - L/2)(|x_0| - L/2)^2 + z_0^2}\ , \end{aligned} \quad (5)$$

where H represents the Heaviside step function. A generalized equation that includes focused imaging is given in the Appendix 6.A. This equation is only valid for points located in the shadow of the line of sensor elements (shaded area in Fig. 1). This equation also has the condition that the minimum transmission delay is zero. The square root term that includes H represents the shortest distance between the virtual source and the transducer (Fig. 2). The H term reduces this distance to $z_0$ if $|x_0| \le L/2$ (Fig. 2, left), otherwise it is the distance to the closest end of the array segment (Fig. 2, right). The limiting value of circular-wave imaging, when $\beta$ tends towards zero, is plane-wave imaging. Substituting (4) into (5) and taking the limit as $\beta \to 0^+$ yields the transmit distance for plane-wave imaging (see appendix 6.B):

$$\lim_{\beta \to 0^+} d_{TX}(X_s) = \left(\mathrm{sgn}(\theta)\frac{L}{2} - x_s\right)\sin(\theta) + z_s\cos(\theta). \quad (6)$$

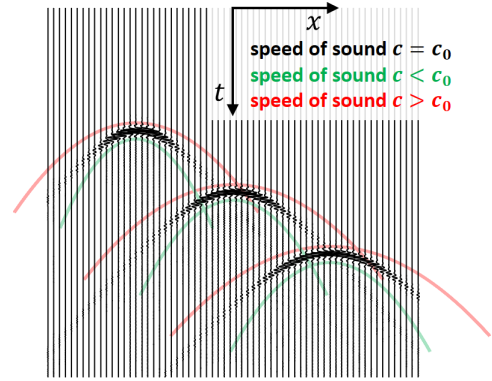

Fig. 3. Hyperbolic signatures in RF signals in the presence of three scatterers. They have an eccentricity of $\approx c$. The hyperbolas flatten as the speed of sound increases. The RF signals were simulated by using SIMUS *[22]*.

## B. Hyperbolic signatures and DAS

The analog signals returned by the piezoelectric elements are commonly referred to as RF (radio frequency) signals. They are amplified and sampled before being processed by a beamformer. We call $s(x_i, t)$ the digital signal recorded by the element #$i$. The time input $t$ is sometimes called fast-time. The argument $x_i$ represents the (lateral) $x$-position of element #$i$ given by (1). In the following, we will define $s_i(t) \equiv s(x_i, t)$. These signals $s_i$ are bandpass modulated signals. The complex envelope of $s_i$, which can be obtained by I/Q demodulation (downmixing + low-pass filtering; see 6.B in the appendix for a short Matlab script), is noted $IQ_i$. The signal $IQ_i(t)$ is complex, with the real and imaginary parts ($I_i(t)$ and $Q_i(t)$) being the in-phase and quadrature components, respectively. Its modulus $|IQ_i(t)| = \sqrt{I_i^2(t) + Q_i^2(t)}$ is the real envelope. A B-mode ultrasound image is obtained by log-compressing real envelopes. Additional post-processing is generally used to enhance the images (e.g. speckle filtering [23]).

When the signals $s_i(t)$ recorded by the array elements ($i = 1 \dots N_e$) are stacked side by side in a $xt$-plane ($x$ − fast time plane), the presence of a scatterer is indicated by a hyperbolic signature (Fig. 3). The equation of the hyperbola $\mathcal{H}_s$ related to a given scatterer of coordinate $X_s = (x_s, z_s)$ can be derived by combining (2) and (3) with $t = \tau_i(X_s) + t_0$ and $x = x_i$:

$$\mathcal{H}_s(X_s) = \left\{(x, t) \in \mathbb{R} \times \mathbb{R}_+ \left| \frac{\left(t - \frac{d_{TX}(X_s)}{c}\right)^2}{z_s^2/c^2} - \frac{(x - x_s)^2}{z_s^2} - 1 = 0\right.\right\}. \quad (7)$$

All the hyperbolas are geometrically similar, i.e. one hyperbola can be rescaled and relocated to coincide with another. Their eccentricity is

$$e = \sqrt{1 + c^2} \approx c. \tag{8}$$

Eq. (8) shows that the higher the speed of sound, the flatter the hyperbolas are (Fig. 3). DAS is a simplified inverse problem that consists in determining an amplitude image in the $xz$-plane from a set of ultrasonic signals of the $xt$-plane. Assuming that the medium consists of a very large number of randomly distributed point scatterers, and neglecting multiple scattering, DAS produces an amplitude image as if snapshotting the contribution of each reflector emitting simultaneously. This is achieved by adding the amplitudes along the hyperbolas $\mathcal{H}_s$ of the $xt$-plane described by (7). The DAS-beamformed signal value at $\boldsymbol{X}_s = (x_s, z_s)$ is written as

$$s_{\mathrm{bf}}(\boldsymbol{X}_s) = \sum_{i \subseteq [\![1, N_e]\!]} s_i(\tau_i(\boldsymbol{X}_s)). \tag{9}$$

The subscript "bf" stands for "beamformed". As we will see later, a subset of $[\![1, N_e]\!]$ may be preferred to consider the element directivity (by using an $f$-number). In other words, it is often advantageous to discard some signals (i.e. use only some $i$ in $1 \ldots N_e$) when using Eq. (9).

Because the signals are obtained by sampling, the term $s_i(\tau_i(\boldsymbol{X}_s))$ are generally unknown and must be estimated from the closest discrete values by using, for example, a $q$-point interpolation; e.g. $q = 1$ for nearest-neighbor interpolation, $q = 2$ for linear interpolation, ... $q = 6$ for 5-lobe Lanczos (windowed sinc) interpolation [24]. As a recall, a $q$-point interpolator can be written as a weighted arithmetic mean of $q$ data points. These interpolating weights can be conveniently included in a beamforming sparse matrix, as explained in the section 2.F entitled "*DAS as a matrix product*".

### C. Beamforming I/Q signals

Note that Eq. (9) is valid **only** for RF signals. In many situations, however, the signals $s_i$ that the elements record are digitally I/Q demodulated before beamforming. I/Q signals are indeed low-frequency signals and are thus easier to handle than RF. More importantly, the magnitudes of the I and Q signals contain amplitude and phase information that are classically used to generate B-mode or Doppler images [25]. I/Q signals can therefore be beamformed directly onto the image grid (e.g. 256×256 size), which cannot be done with RF signals in general. The beamforming of RF signals indeed requires fine axial sampling to ensure correct envelope detection.

Summation and demodulation are non-commuting operators. To preserve the relative phases when delaying-and-summing I/Q signals after a demodulation, phase rotators are necessary (e.g. see Eq. 4 in [26]). Beamformed I/Q signals can thus be written as:

$$IQ_{\mathrm{bf}}(\boldsymbol{X}_s) = \sum_{i \subseteq [\![1, N_e]\!]} IQ_i(\tau_i(\boldsymbol{X}_s)) \, e^{2i\pi f_c \, \tau_i(\boldsymbol{X}_s)}. \tag{10}$$

The center frequency $f_c$ is also the frequency used for downmixing during I/Q demodulation. Once beamformed I/Q ($IQ_{\mathrm{bf}}$) are obtained: 1) a B-mode image can be generated by taking their log-compressed moduli, or 2) a Doppler image can be produced by analyzing the temporal phase shifts from one frame to another. Only a few correctly selected $IQ_i$ need to be summed along the hyperbolas in Eq. (10) to obtain beamformed data that can yield high-quality images. We will now see how to select an appropriate subset of $[\![1, N_e]\!]$, i.e. how to choose a proper receive aperture.

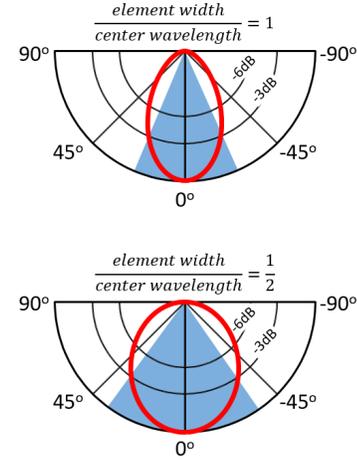

Fig. 4. The directivity (Eq. 11) of one array element (thick closed curve) depends on the width of the element in relation to the wavelength. The conical sectors show areas where the directivity is greater than -3dB.

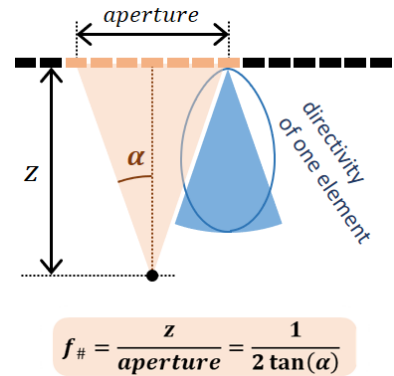

Fig. 5. An appropriate $f$-number $f_\#$ can be determined from the directivity of the array elements. The $f$-number is related to the so-called angle-of-view ($= 2\alpha$). The directivity curve and conical sector are those of Fig. 4.

### D. Receive $f$-number

The signal amplitude along a hyperbolic signature $\mathcal{H}_s$ is not uniform. It is maximal at its vertex whose abscissa is $x = x_s$. What essentially governs the signal amplitude along the hyperbolas is the directivity of the elements. The elements are indeed not omnidirectional. As a result, they do not receive uniformly in all directions. If the ratio ($W/\lambda$) between the element width ($W$) and the wavelength ($\lambda$) is relatively large, the element has a high directivity: it receives mainly in front of it. The smaller this ratio is, the more omnidirectional it becomes. This ratio is ~1 in most vascular linear arrays and ~0.5 in most cardiac phased arrays (note that in clinical practice, cardiac phased arrays for transthoracic echocardiography are uniform linear arrays). The directivity of an element depends on the wave-propagation angle $\vartheta \in \left] \frac{-\pi}{2}, \frac{\pi}{2} \right[$ with respect to the $z$-axis (Fig. 4). For a piston-like element in a soft baffle [27], it is given by

$$D(\vartheta) = \cos \vartheta \, \operatorname{sinc}\left( \pi \frac{W}{\lambda} \sin \vartheta \right). \tag{11}$$

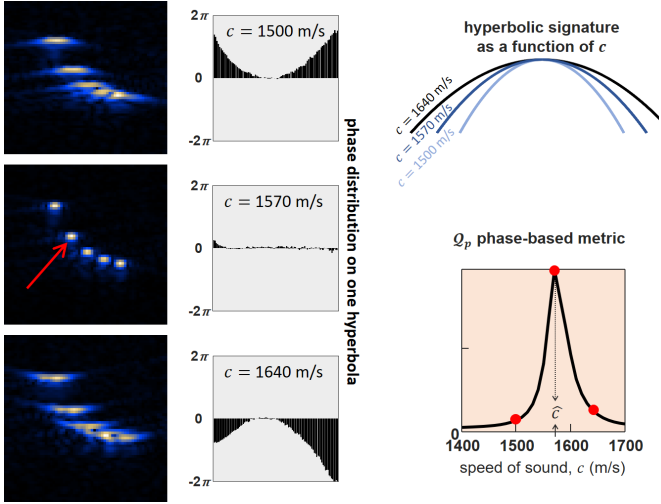

Fig. 6. Effect of the speed of sound on DAS beamforming. Inadequate speeds can lead to distorted (gamma-compressed) B-mode images (left column) because the received signals are summed on erroneous hyperbolic paths (top right). The proposed $\mathcal{Q}_p$ metric analyzes the phase distribution on these hyperbolas (center column). An appropriate speed of sound tends to uniformize the distribution of the phase and maximizes $\mathcal{Q}_p$ (bottom right).

The sound pressure received by an element is directly proportional to $D(\vartheta)$ and thus has a maximum amplitude when $\vartheta = 0$. It decreases as $\vartheta$ tends towards $\pm \pi/2$ (Fig. 3). It follows that the signal-to-noise ratio along a hyperbola $\mathcal{H}_s$ also decreases when $|x - x_s|$ increases, i.e. when one moves away from the vertex (Fig. 3). For this reason, it is recommended to use the top part only of the hyperbolas during receive beamforming by using an $f$-number. The receive $f$-number is defined by the ratio between the depth $z_s$ and the width of the receive aperture. It is related to the angle-of-view $2\alpha$ (defined in Fig. 5) as follows:

$$f_\# = \frac{1}{2 \tan(\alpha)}. \tag{12}$$

By using the directivity expression (11), the angle-of-view should satisfy:

$$D(\alpha) = D_{thresh} \implies$$
$$\alpha = \left\{ \vartheta \in \left[0, \frac{\pi}{2}\right] \mid \cos \vartheta \, \operatorname{sinc}\left(\pi \frac{W}{\lambda} \sin \vartheta\right) = D_{thresh} \right\} \tag{13}$$

On the basis of image quality and SNR (signal-to-noise ratio), a rule-of-thumb compromise is to discard amplitudes below values of -3dB (i.e. $D_{thresh} = 0.71$). If $W/\lambda = 1$, the expressions (12) and (13) yield $f_\# = 1.2$ (see 6.B in the appendix for a Matlab script). As shown by Eq. (11), the element directivity is wavelength-dependent. Consequently, it is preferable to consider the smallest significant wavelength of the signals. For a signal of center frequency $f_c$ and bandwidth $B$ (both in Hz), the

angle-of-view $\alpha$ in Eq. (13) can thus be determined by using:

$$\lambda = \lambda_{\min} = c/f_{\max} = c/(f_c + B/2). \tag{14}$$

Taking into account the $f$-number, the DAS equation (10) at $X_s = (x_s, z_s)$ then becomes:

$$IQ_{\mathrm{bf}}(X_s) = \sum_{i \in [\![1, N_e]\!]} IQ_i\big(\tau_i(X_s)\big) \, e^{2 i \pi \, f_c \, \tau_i(X_s)},$$
$$\text{with } i \text{ subject to } \frac{z_s}{2 \, |x_s - x_i|} \geq f_\#. \tag{15}$$

### E. Speed of sound

As shown by Eq. (8), the eccentricity of the hyperbolas is almost equal to the speed of sound $c$. In other words, the speed of sound governs the shape of the hyperbolas. If the $c$ parameter in DAS does not match the actual speed of sound, the DAS algorithm adds the signal amplitudes along incorrect hyperbolas, which distorts the output point spread function. It follows that an error on $c$ may affect image quality significantly (Fig. 6). It is generally assumed that the speed of sound in soft tissues is 1540 m/s on average. *In vivo*, this may be untrue if the medium is relatively heterogeneous [28]. *In vitro*, temperature or storage conditions can modify the mechanical properties of a phantom and therefore the speed of sound [29]. To complicate matters, several factors might tend to reduce the average speed of sound: fat tissues, presence of bubbles in the coupling gel, air interface, curved or bent rays due to aberrations. If the average speed of sound in the insonified medium is not known, the hyperbolas used during DAS may deviate significantly from the actual hyperbolas (Fig. 6). As the signal phase is not uniform along incorrect hyperbolas, this inaccuracy can have negative effects on the quality of ultrasound images. We propose one solution to estimate an optimal average speed of sound. The idea is to determine the speed of sound that returns the hyperbolas with minimal phase dispersion. The technique that we introduce is inspired by Yoon *et al.* [30]. We define the following dimensionless phase-based quality metric for a given speed of sound:

$$\mathcal{Q}_p(c) = \sum_{k = 1 \dots M} \frac{\left| IQ_{\mathrm{bf}}(X_s(k)) \right|^2}{\operatorname{Var}(\varphi)_k}. \tag{16}$$

$M$ is the number of spatial points where the signals are beamformed. The numerator represents the squared real envelope (intensity) of the signal, after DAS beamforming, at point $X_s(k)$. The denominator is the sample variance of the unwrapped phase along the $k^{\text{th}}$ hyperbola $\mathcal{H}_s$ associated with the scatterer located at $X_s(k)$. Note that $\mathcal{Q}_p$ can also be calculated from compound data to return better estimates. In this case, the variances are estimated on compound hyperbolas obtained by coherent summation. We define the expected speed of sound $\hat{c}$ as the one that uniformizes the phases along the hyperbolas, i.e. $\hat{c}$ maximizes the $\mathcal{Q}_p$ metric (Fig. 6):

$$\hat{c} = \arg \max_c \big( \mathcal{Q}_p(c) \big). \tag{17}$$

In [30], only the sum of the phase variances – the denominator in (16) – was minimized, without taking into account the respective intensities. The $Q_p$ metric considers these intensities – the numerator in (16) – to prioritize the contribution of the bright speckles. The appendix 6.D provides a simplified Matlab function (called *ezsos*) to estimate the speed of sound by maximizing (16). Note that we propose to estimate an average speed of sound (i.e. a constant), not a mapping. The region of interest, and its average speed of sound, remain more or less time-invariant when scanning an organ. A single transmit is thus needed. The I/Q signals can be beamformed at a limited number of points (say a few hundred). This process has to be done once.

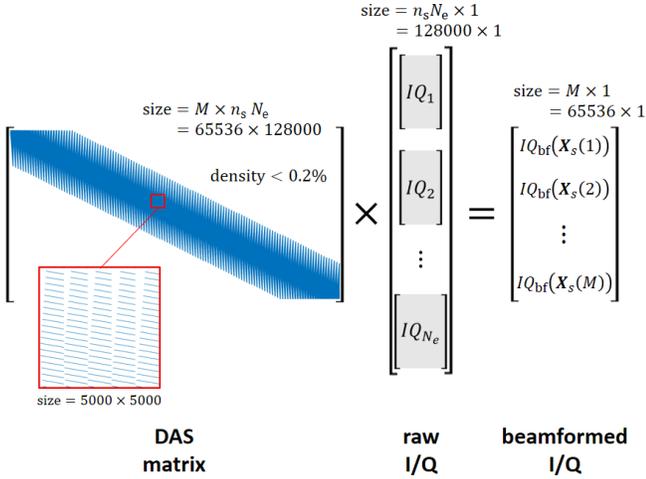

Fig. 7. Example of a sparse DAS matrix. DAS beamforming can be written as a sparse matrix-vector multiplication. For example, the generation of a 256×256 image by beamforming 128 signals each containing 1000 samples leads to a DAS matrix of size (256×256) × (128×1000) = 65536 × 128000. Note that the matrix shown is for illustration purpose only (its aspect can be quite different). $M$ = number of beamformed data (is generally chosen equal to the number of pixels of the image to be formed); $n_s$ = number of time samples per raw signals; $N_e$ = number of array elements.

### F. DAS as a matrix product

As can be seen by Eq. (15), the DAS is a linear operator that transforms the I/Q signals recorded by the array elements ($IQ_i$ with $i = 1 \dots N_e$) to beamformed I/Q data ($IQ_{bf}$). If the time series $IQ_i$ contain $n_s$ samples each, and are stacked in a **column** vector $\boldsymbol{IQ} = \left[IQ_1, \dots, IQ_{N_e}\right]^{\mathrm{T}}$ of size ($n_s N_e \times 1$), DAS beamforming can be written synthetically as:

$$IQ_{bf} = \mathbf{M}_{DAS} \times IQ, \tag{18}$$

where $\mathbf{M}_{DAS}$ is the DAS matrix (Fig. 7) that contains the interpolation weights (see the end of paragraph 2.B), the phase rotators (see equation 10), and the sum truncation induced by a positive f-number (see equation 15). If the data (e.g. B-mode or Doppler) to be constructed contain $M$ pixels, then the column vector $IQ_{bf}$ that contains the beamformed I/Q signals is of length $M$. The matrix $\mathbf{M}_{DAS}$ is of size ($M \times n_s N_e$). It is complex due to the phase rotations present in Eq. (15). The matrix-vector product for DAS beamforming is schematized in Fig. 7. It is understood that RF signals can also be processed by this matrix product, in which case the DAS matrix is real. Nevertheless, sufficiently fine axial sampling is required to allow subsequent envelop detection. For this reason, unless it is strictly necessary to post-process RF signals (e.g. in RF-based motion tracking [31]), this approach is not recommended as it unnecessarily burdens data storage and calculations. As mentioned earlier, fine axial sampling is indeed required to allow subsequent envelope detection in beamformed RF signals (e.g. by demodulation or through a Hilbert transform). This is not the case when beamforming I/Q. In an extreme situation, it is possible to beamform I/Q onto a single pixel whereas it is not with RF (how to detect the envelope with a single RF spot?). In short, the original I/Q can be easily decimated to reduce their number of samples ($n_s$), and they require far fewer beamforming points ($= M$). This significantly reduces the number of computational steps (proportional to $M \times n_s$, Fig. 7) when DASing.

It is to be noted that several frames (all with the same virtual source) can be beamformed simultaneously [32]. In such case, $\boldsymbol{IQ}$ becomes a matrix whose each column contains the $n_s N_e$ I/Q data values of each frame. Now, to obtain the interpolated I/Q along the hyperbolas required for estimating the metric $Q_p$ (16), the signals $IQ_{i, \ i=1 \dots N_e}$ must be arranged in a matrix of size ($n_s N_e \times N_e$), as represented in Fig. 14 of the appendix (see also the Matlab function *ezsos* in the appendix 6.C).

$\mathbf{M}_{DAS}$ is large but very sparse since it mostly contains zero-valued elements (Fig. 7). Let $nnz$ denote its number of nonzero elements. If a $q$-point interpolation is used to estimate the signals at $\tau_i(\boldsymbol{X_s})$, we have $nnz \leq (MN_eq)$. The right-hand term is an upper bound since limited apertures (as $f_\# > 0$) can be used during receive beamforming, which reduces $nnz$. The sparsity [33] of the DAS matrix verifies:

$$sparsity(\mathbf{M}_{DAS}) = 1 - \frac{nnz}{Mn_sN_e} \geq 1 - \frac{q}{n_s}. \tag{19}$$

To give an example, if each array element acquires 1000 I/Q samples and if the signals $IQ_i$ are interpolated linearly (i.e. $q = 2$), then the DAS matrix has a sparsity greater than 1-2/1000 = 99.8%, i.e. a density smaller than 0.2%. Sparse matrix-vector multiplication (SpMV) can be computed on GPUs [34]. Its computational complexity is $O(nnz) \leq O(MN_eq)$ [33]. A construction of the complex DAS matrix (for I/Q beamforming) is given in the Matlab function (called *ezdas*) included in the appendix. The proposed function *ezdas* uses a linear interpolation, which is recommended in most situations. For the sake of completeness, the *dasmtx* function of the MUST toolbox (www.biomecardio.com/MUST/) includes several interpolation methods: nearest neighbor, linear, quadratic, 5-point least-squares parabolic, and 3-or-5-lobe Lanczos.

Fig. 7 illustrates the sparsity of such a matrix. If the ultrasound sequence (array, transmit, receive) and the $M$ beamforming point locations remain unchanged, $\mathbf{M}_{DAS}$ only needs to be calculated once. Once the DAS matrix is created (or loaded), SpMV-based beamforming is very fast and compatible with real-time visualization [32].

## 3. In vitro and in vivo examples

The respective effects of the speed of sound and $f$-number on image quality were evaluated using the PICMUS dataset available online [35]. We analyzed RF data that were acquired in an *in vitro* CIRS phantom (040GSE) and sampled at four times the center frequency. These data were obtained by transmitting steered plane waves with a 128-element 5-MHz linear array. The RF data were IQ-demodulated then beamformed using the DAS matrix of Eq. (18) on the Cartesian image grid specified in PICMUS. A series of real envelopes were constructed from one unsteered plane wave (no transmission delay) for a large range of speed of sound (1400 to 1700 m/s) or $f$-number (0 to 4). For reference, high-quality real envelopes were also generated by coherent compounding with the whole dataset (75 plane waves with steering ranging from -16 to +16 degrees). The optimal $f$-number was estimated by using Eq. (12) and (13) with $D_{thresh} = 0.71$ (-3dB threshold). The optimal speed of sound was determined by maximizing the phase-based metric using Eq. (16). Contrast and lateral resolution were quantified using the contrast-to-noise ratio (CNR) and full width at half maximum (FWHM), respectively, as detailed in [35]. The CNR and FWHM reported in our work (Fig. 9) were an average from two cysts and seven nylon wires, respectively (see Figures 2b and 2d in [35]).

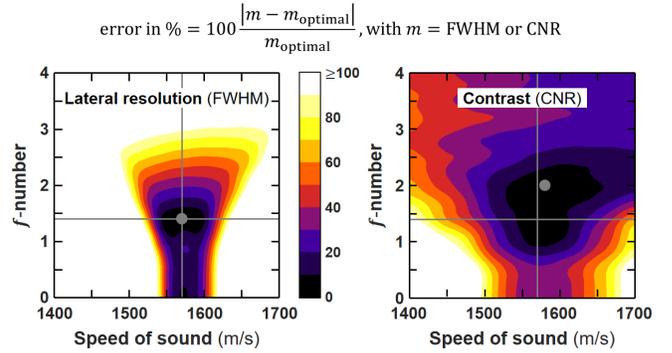

error in % = $100 \cdot \frac{|m - m_{optimal}|}{m_{optimal}}$, with $m$ = FWHM or CNR

Fig. 9. *In vitro* results obtained in a CIRS phantom using single plane wave imaging: effect of the speed of sound and $f$-number on FWHM and CNR. The errors in % are expressed relative to the optimum values (shown by the dot). The horizontal and vertical lines represent the optimal speed of sound and $f$-number estimated by the equations introduced in the manuscript.

### A. In vitro results

From Eq. (14), the minimum significant wavelength transmitted by the elements was $\lambda_{min} \approx 0.23$ mm. It followed that $W/\lambda_{min} \approx 1.17$. From Eq. (12) and (13), with $D_{thresh} = 0.71$ (i.e. -3dB directivity threshold), the directivity-based $f$-number was thus $f_\# \approx 1.4$. Maximizing $\mathcal{Q}_p$ (16) by emitting one plane wave returned an optimal speed of sound $\hat{c} = 1570$ m/s.

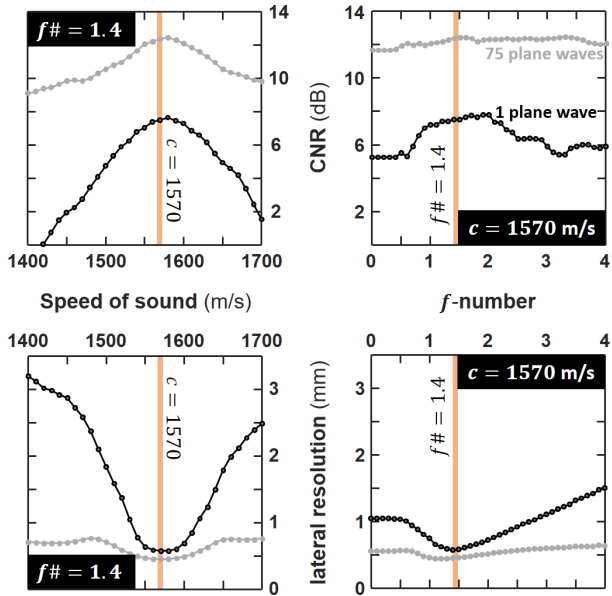

Fig. 8. *In vitro* results obtained in a CIRS phantom (experimental data from PICMUS). The top and bottom rows show the effect of the speed of sound and $f$-number on CNR (contrast-to-noise ratio) and lateral resolution, respectively. The thick vertical bars represent the optimal speed of sound and $f$-number estimated by the equations introduced in the manuscript. The gray dots represent the metrics after compounding 75 plane waves (reference values), while the black dots are for a single plane wave transmit.

The speed of sound was also measured in the carotids of eight healthy volunteers by maximizing the $\mathcal{Q}_p$ metric. A cardiologist used a linear-array transducer (ATL L7-4, center frequency = 5.2 MHz, element width = 0.245 mm, fractional bandwidth at -6dB = 65%) connected to a Verasonics scanner (V-1-128, Verasonics Inc.) to scan the common carotid artery (protocol similar to that described in [36]). Five steered plane waves (transmit beam angles evenly spaced between -10° and 10°)

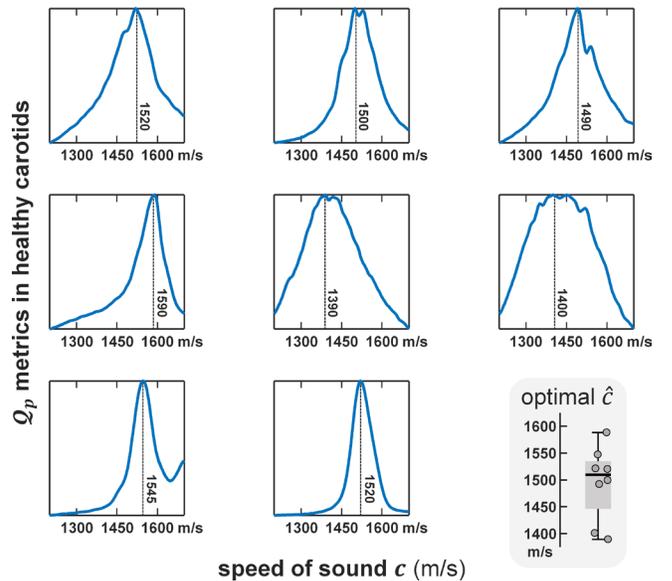

Fig. 10. *In vivo* results obtained in 8 carotids. The curves represent the $\mathcal{Q}_p$ metric as a function of the speed of sound. The boxplot shows the distribution of the estimated speeds of sound (median, range, 25th, 75th percentiles).

As anticipated, both the speed of sound and the $f$-number had a significant impact on the contrast-to-noise ratio (CNR) and lateral resolution (Fig. 8 and Fig. 9). In the CIRS phantom, CNR and lateral resolution were both optimal for $c \approx 1550$- 1600 m/s and $f_\# \approx 1.1$-$1.6$. These values determined using a brute-force search were consistent with those estimated by physical reasoning (phase homogeneity and element directivity) using Eq. (12), (13) and (16), (17); see Fig. 8 and Fig. 9.

### B. In vivo results

The calculated $f$-number was also $f_\# \approx 1.4$ (same transducer as *in vitro*). The speeds of sound estimated through maximizing $Q_p$ in the eight carotids were $1510 \pm 41$ m/s (Fig. 10). Fig. 11 shows a carotid whose speed of sound was estimated at 1400 m/s. This value may seem abnormally low. However, this speed of sound probably does not reflect the average speed of sound in the carotid artery. This value has somewhat improved the image by maximizing the metric we introduced. This apparent speed of sound probably reduced the deterioration of the image caused by a combination of independent effects (fat, air, aberrations...). The $f$-number had an obvious positive impact on image quality (Fig. 11, 1st row vs. 2nd row). In comparison, the effect of the speed of sound was less noticeable, except for the change of scale in the $z$-direction (Fig. 11, 1st column vs. 2nd column).

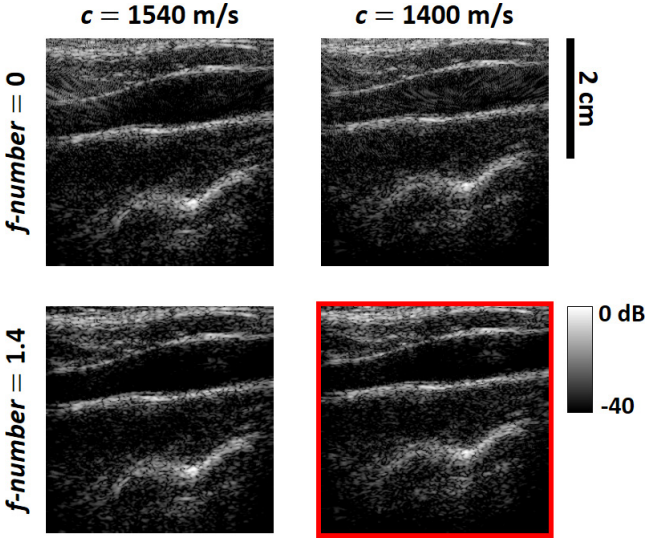

Fig. 11. *In vivo* example in one carotid. The directivity-derived $f$-number was 1.4 and the estimated speed of sound was 1400 m/s (instead of the assumed 1540 m/s). $f$-number = 0 means "full aperture". The artefacts related to a full aperture are well visible in the lumen (top row).

## 4. Discussion

In this essentially educational article, we have considered the fundamentals of the delay-and-sum (DAS) for ultrasound image beamforming. The underlying theory has been described in depth so that users can be aware of the functioning and limitations of this technique. In particular, we have explained that the DAS can be written numerically as a sparse matrix-vector product and we have clarified the choice of the $f$-number and speed of sound from a physical and practical viewpoint.

### A. The $f$-number and speed of sound in DAS

We have seen that the $f$-number and speed of sound both influence image quality returned by DAS (Fig. 8). Although their respective effects are well known, how to set these parameters had not been clearly explained in the literature. We have addressed this issue by proposing two simple approaches that can be summarized by equations (12)-(13) and (16)-(17). As we have argued, for a relatively homogeneous medium, the $f$-number is essentially related to the directivity of the array elements. It depends on the transducer (element width, center frequency) and pulse-echo bandwidth. A modification is needed when angled receiving is performed, as in vector Doppler [36], [37]. In this case, the angle variable $\vartheta$ in Eq. (13) must be replaced by $(\vartheta + |\theta_{RX}|)$, with $\theta_{RX}$ being the receive steering angle.

An appropriately chosen $f$-number helps to optimize the bias-variance tradeoff when summing the signals along the hyperbolas. An $f$-number that is too large (too small an aperture) induces a summation based on a small sample of elements, thus generating a significant bias. In the case of an $f$-number that is too small (too large an aperture), the summation involves low directivities (and therefore low SNRs), thus causing a high variance. Too high bias or variance affects the contrast and lateral resolution (Fig. 8). We heuristically found that a -3dB-directivity threshold gives a good compromise (i.e. Eq. (13) with $D_{thresh} = 0.71$). Similar conclusions could be reached with Doppler or vector Doppler. Fig. 12 illustrates one vector Doppler example recomputed from [36] (plane wave imaging, receive angle $\theta_{RX} = \pm 15°$, 3 cm-diameter disk rotating at 300 rpm). The smallest velocity-vector errors coincided with the directivity-derived $f$-number ($\approx 2.6$) obtained from Eq. (13) with $(\vartheta + |\theta_{RX}|)$ instead of $\vartheta$.

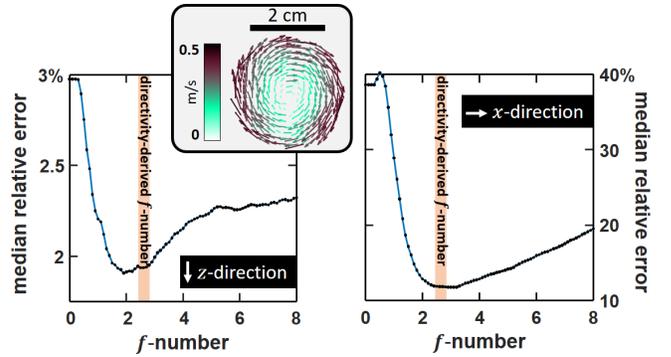

Fig. 12. Vector Doppler obtained in a rotating disk (data from [36]): 17 unsteered plane waves were transmitted; I/Q signals were DAS-beamformed with $\pm 15°$ receive angles. The curves show the median relative errors on the $x$- and $z$-velocity components as a function of the $f$-number. The thick vertical bars represent the optimal $f$-number estimated by Eq. (12)-(13).

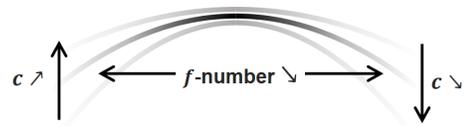

Fig. 13. Knowing the speed of sound ($c$) allows one to determine which hyperbola must be chosen. The $f$-number allows one to decide which part of the hyperbola should be considered.

To put it simply, the $f$-number allows one to decide which part of the hyperbola should be considered. Knowing the speed of sound, on the other hand, allows one to determine which hyperbola should be chosen (Fig. 13). According to the *in vitro* results obtained in a calibration phantom, the speed of sound has a significant impact on the standard CNR and FWHM metrics (Fig. 8). However, when considering the *in vivo* results, the overall physiological appearance appears almost unchanged (Fig. 11), except for the axial dimensions. This would probably have little or no diagnostic impact in a clinical setting. It then seems that adjusting the speed of sound would be impactful mainly for calibration or *in vitro* comparisons. Although this is only a speculation, there might nevertheless be some interest in super-resolution localization of microbubbles, as each microbubble is expected to be located at the peak or centroid of the bubble-PSF (point spread function) [38]. Optimizing the speed of sound and thus refining the PSF could make localization more robust. On a more clinical level, it has been shown that the longitudinal speed of sound has a potential diagnostic value in hepatic steatosis, a common liver disease [39], [40]. In particular, Imbault *et al.* [39] estimated the speed of sound by locally analyzing the spatial coherence [41]. The measurement of the proposed dimensionless metric $Q_p$ (16) based on I/Q signals could provide an alternative method.

### B. DAS vs. advanced beamformers

As stated earlier, DAS is the best-known and simplest beamforming technique in medical ultrasound imaging. It is fast, easy to program and well efficient in most situations. Moreover, DAS allows for low-complexity real-time beamforming [42]. Yet, several studies have focused on adaptive beamformers whose intended purpose is to improve image quality for medical ultrasound. One typical approach for adaptive beamforming is to perform a weighted sum and adjust the weights to the data. This is similar to an ordinary DAS, except that instead of each of the signal samples (inside the receive aperture) contributing equally to the final sum, some data samples contribute more than others do, and the relative contributions depend on these samples. Minimum variance (MV) and its variant (eigen-based MV) as well as *p*-DAS are ones of these techniques [43]–[45]. Adaptive methods based on the coherence factor [46] also belong in this category, though the weights are pixelwise. Because they are data-dependent, some adaptive beamforming techniques are computationally expensive and cannot be used in real-time. More importantly, although they can improve common metrics (e.g. CNR, FWHM) in specific anechoic-cyst- or wire-based calibration phantoms, it is not clear if they have significant value in a clinical context, as to whether they can help in a better diagnosis at the patient's bedside. Another downside of this adaptability is that local phase information, from one emission to the next, can be distorted. This may have a detrimental impact on velocity estimation by color Doppler: the more data-dependent the beamformer, the higher the Doppler error [47]. To avoid biased conclusions when comparing alternative beamformers against DAS through these metrics, it is therefore essential to ensure that the speed of sound and/or the *f*-number have been correctly set for DAS beamforming.

Another increasingly growing direction for ultrasound image formation is deep learning. Deep learning (DL) can intervene at different stages, for example at the beamforming [48] or compounding [49] level. If properly trained, DL systems are expected to provide high-quality images with fewer input data. Data-driven DL algorithms, however, are highly dependent on accurate and clean training datasets to learn from: for DL systems to learn to produce high-quality images, they must be heavily trained with high-quality images. The question then arises as to how to provide such images of optimal quality for a wide variety of organs. If DAS is the chosen beamformer to train DL systems, it is essential to ensure that its parameters are correctly set.

## 5. Conclusion

DAS is highly popular in ultrasound imaging. If properly designed and compound, it can be very effective in terms of *in vivo* image quality, even in challenging situations such as the beating heart *[50]*. Yet DAS is sometimes considered mediocre by investigations based mainly on *in vitro* experiments because it is often suboptimally addressed. This opinion is often biased by its misuse. For example, the *f*-number is sometimes neglected or wrongly chosen. It is imperative that the receive *f*-number be properly selected. In addition, the choice of the speed of sound can have a significant impact on the metrics measured *in vitro*. To get the most out of DAS and to prevent any misguided discredit, it is suggested to optimize its parameters in order to use it wisely. When it comes to DAS, one must not DAS out of step.

## 6. Appendix

### A. Generalized transmit distance (focused and circular beams)

If $(x_0, z_0)$ is a focus point, a reasoning similar to that used to derive Eq. (5) leads to:

$$d_{\text{TX}}(X_s) = \text{sign}(z_s - z_0)\sqrt{(x_s - x_0)^2 + (z_s - z_0)^2} + \sqrt{H(|x_0| + L/2\,)(|x_0| + L/2)^2 + z_0^2}\,. \tag{20}$$

Equations (5) and (20) can be combined to yield a generalized transmit distance (focusing or diverging wavefronts):

$$d_{\text{TX}}(X_s) = \text{sign}(z_s - z_0)\sqrt{(x_s - x_0)^2 + (z_s - z_0)^2} \tag{21}$$
$$+ \text{sign}(z_0)\sqrt{H(|x_0| + \text{sign}(z_0)\,L/2\,)(|x_0| + \text{sign}(z_0)\,L/2)^2 + z_0^2}\,.$$

### B. Plane wave as a limit of diverging wave

Taking the limit of the virtual source coordinates, given by Eq. (4), as $\beta \to 0^+$ yields

$$\begin{cases} \lim\limits_{\beta \to 0^+} x_0 = \dfrac{L}{2}\,\dfrac{\sin(2\theta)}{\beta} = L\,\dfrac{\cos(\theta)\sin(\theta)}{\beta}, \\[2mm] \lim\limits_{\beta \to 0^+} z_0 = -\dfrac{L}{2}\,\dfrac{1 + \cos(2\theta)}{\beta} = -L\,\dfrac{\cos^2(\theta)}{\beta}, \end{cases} \tag{22}$$

Let $f(\theta) = L\cos(\theta)\sin(\theta)$ and $g(\theta) = L\cos^2(\theta)$. The transmit distance (5) as $\beta \to 0^+$ is thus reduced to

$$\begin{aligned} &\lim_{\beta \to 0^+} d_{\text{TX}}(x_s, z_s) \\ &= \sqrt{(x_s - f(\theta)/\beta)^2 + (z_s + g(\theta)/\beta)^2} \\ &- \sqrt{(\text{sgn}(\theta)\,f(\theta)/\beta - L/2)^2 + (g(\theta)/\beta)^2}\,. \end{aligned} \tag{23}$$

Expanding the four square terms then factoring yields:

$$\lim_{\beta \to 0^+} d_{\text{TX}}(x_s, z_s)$$
$$= \frac{\sqrt{f(\theta)^2 + g(\theta)^2}}{\beta}$$

$$\left[ \sqrt{1 + \frac{2(z_s g(\theta) - x_s f(\theta))}{f(\theta)^2 + g(\theta)^2} \beta + \frac{x_s^2 + z_s^2}{f(\theta)^2 + g(\theta)^2} \beta^2} \right. \tag{24}$$

$$\left. - \sqrt{1 - \frac{\text{sgn}(\theta) f(\theta) L}{f(\theta)^2 + g(\theta)^2} \beta + \frac{L^2}{4(f(\theta)^2 + g(\theta)^2)} \beta^2} \right].$$

Using the Taylor series of $(1 + x)^{1/2}$ at the first order yields:

$$\lim_{\beta \to 0^+} d_{\text{TX}}(x_s, z_s) = \frac{f(\theta)(\text{sgn}(\theta) L - 2x_s) + 2g(\theta) z_s}{2\sqrt{f(\theta)^2 + g(\theta)^2}} \tag{25}$$

Noting that $\cos(\theta)$ is always positive (since $\theta \in \,]\frac{-\pi}{2}, \frac{\pi}{2}[$), Eq. (25) can be simplified to obtain Eq. (6) after replacing $f(\theta)$ and $g(\theta)$.

### C. Determine the f-number in Matlab

```
% Note: in Matlab, sinc(x) = sin(pi*x)/(pi*x)
f = @(th,width,lambda)...
    abs(cos(th)*sinc(width/lambda*sin(th))-0.71);
alpha = fminbnd(@(th) f(th,width,lambda),0,pi/2);
fnumber = 1/2/tan(alpha);
```

### D. A simple Matlab code for DAS beamforming

```
function [bfSIG,M] = ezdas(SIG,x,z,vsource,param)

%EZDAS    Delay-And-Sum beamforming
%        (Easy version of DAS)
%
%   BFSIG = EZDAS(SIG,X,Z,VSOURCE,PARAM) beamforms
%   the RF or I/Q signals stored in the array SIG,
%   and returns the beamformed signals BFSIG. The
%   signals are beamformed at the points specified
%   by X and Z.
%
%   1) SIG must be a 2-D array. The first dimension
%      (i.e. each column) corresponds to single RF
%      or I/Q signals over (fast-) time, with the
%      1st column corresponding to the 1st element.
%   2) VSOURCE contains the coordinates [x0,z0] of
%      the virtual point source. Use large x0,z0 for
%      plane waves.
%   3) PARAM is a structure that contains the
%      parameter values required for the delay-and-
%      sum (see below for details).
%
%   Note: SIG must be complex for I/Q data
%         (i.e. SIG = complex(I,Q) = I + 1i*Q).
%
%   [~,M] = EZDAS(...) also returns the DAS matrix.
%
%   PARAM must contain the following fields:
%   ---------------------------------------
%   1) PARAM.fs: sampling frequency (Hz)
%   2) PARAM.pitch: element pitch (m)
%   3) PARAM.fc: center frequency (Hz)
%   4) PARAM.c: longitudinal velocity (m/s)
%   5) PARAM.fnumber: receive f-number
%
%   ---
%   NOTE #1: Interpolation method: EZDAS uses a
%   linear interpolation to generate the DAS matrix.
%   ---
%   NOTE #2: EZDAS is for pedagogical purpose. Use
%   DAS of the MUST toolbox for more options.
%   ---
%
%   See also DAS, DASMTX.
%
%   -- Damien Garcia -- 2019/11
%   www.biomecardio.com

siz0 = size(x);
[nl,nc] = size(SIG);
x = x(:); z = z(:);

% ULA (uniform linear array):
% x-coordinates of the elements
xe = ((0:nc-1)-(nc-1)/2)*param.pitch;
L = xe(end)-xe(1); % length of the array

% coordinates of the virtual source
x0 = vsource(1); z0 = vsource(2);

% transmit & receive distances
dTX = hypot(x-x0,z-z0)-...
    hypot((abs(x0)-L/2)*(abs(x0)>L/2),z0);
dRX = hypot(x-xe,z);

% two-way travel times
tau = (dTX+dRX)/param.c;

% fast-time indices
idxt = tau*param.fs + 1;

% boolean vectors
I = idxt>=1 & idxt<=nl-1;
Iaperture = abs(x-xe) <= (z/2/param.fnumber);
I = I&Iaperture;

% linear indices
idx = idxt + (0:nc-1)*nl;
idx = idx(I);
idxf = floor(idx);
idx = idxf-idx;

% DAS matrix
[i,~] = find(I);
s = [idx+1;-idx]; % (for linear interpolation)
if ~isreal(SIG) % if IQ: phase rotations
    s = s.*exp(2i*pi*param.fc*[tau(I);tau(I)]);
end
M = sparse([i;i],[idxf;idxf+1],s,numel(x),nl*nc);

% DAS beamforming
bfSIG = reshape(M*SIG(:),siz0);
```

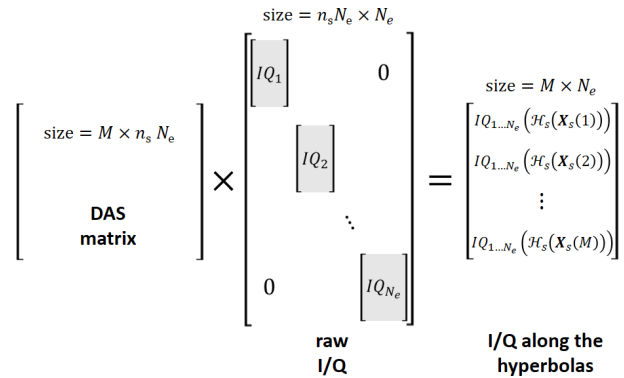

**Fig. 14.** DAS matrix product to obtain interpolated I/Q along the $M$ hyperbolas $\mathcal{H}_s$ associated to the $M$ pixels of the beamformed image. $n_s$ = number of time samples per raw signals; $N_e$ = number of array elements.

### E. A simple Matlab code for estimating the speed of sound

```
function c = ezsos(IQ,x,z,vsource,param)
```

```
%EZSOS   Speed-of-sound estimation
%   (Easy version of SOS)
%
%   c = EZSOS(IQ,X,Z,VSOURCE,PARAM) returns the
%   speed of sound that yields an "optimal" real-
%   envelope image by DAS beamforming. The optimal
%   speed of sound is estimated by analyzing the
%   phases along the diffraction hyperbolas whose
%   vertices are located at (X,Z).
%
%   The input arguments are the same as those of
%   EZDAS. Except that the signals MUST be complex
%   (I/Q data). Enter "help EZDAS" for details.
%
%   See also SOS, EZDAS.
%
%   -- Damien Garcia & Vincent Perrot -- 2019/11
%   www.biomecardio.com

assert(~isreal(IQ))
[nl,nc] = size(IQ);
c0 = param.c;

IQ0 = IQ;
IQ = sparse(1:nl*nc,kron(1:nc,ones(1,nl)),...
    IQ(:),nl*nc,nc);

c = round(fminbnd(@PBC,1200,1700));

    function Qp = PBC(c)
        param.c = c;
        [~,M] = ezdas(IQ0,x,z/c0*c,vsource,param);

        hyperb = full(M*IQ);
        % (contains the diffraction hyperbolas)

        A = angle(hyperb);
        A(A==0) = NaN;
        A = unwrap(A,[],2);
        Qp = abs(mean(hyperb,2,'omitnan')).^2./...
            std(A,0,2,'omitnan').^2;
        Qp = -mean(Qp,'omitnan');

    end

end
```

### F. A simple Matlab code for I/Q demodulation

```
function IQ = ezrf2iq(RF,Fs,Fc)

%EZRF2IQ   I/Q demodulation of RF data
%   (easy version of RF2IQ)
%
%   IQ = RF2IQ(RF,Fs,Fc) demodulates the radio-
%   frequency (RF) bandpass signals and returns the
%   Inphase/Quadrature (I/Q) components. IQ is a
%   complex whose real (imaginary) part contains the
%   inphase (quadrature) component.
%
%   1) Fs is the sampling frequency (in Hz),
%   2) Fc represents the center frequency (in Hz).
%
%   EZRF2IQ demodulates along columns for 2-D and
%   3-D RF data. Each column corresponds to a single
%   RF signal over (fast-) time.
%
%   EZRF2IQ is for pedagogical purpose. You may use
%   RF2IQ for more options.
%
%   See also RF2IQ.
%
%   -- Damien Garcia -- 2019/11
%   www.biomecardio.com

%-- Convert to column vector (if RF is a row vector)
wasrow = isrow(RF);
```

```
if wasrow, RF = RF(:); end

%-- Time vector
nl = size(RF,1);
t = (0:nl-1)'/Fs;

%-- Downmixing of the RF signals
IQ = double(RF).*exp(-1i*2*pi*Fc*t);

%-- Low-pass filter
[b,a] = butter(5,.5);
IQ = filtfilt(b,a,IQ)*2;

%-- Recover the initial size (if was a row vector)
if wasrow, IQ = IQ.'; end
```

## References

[1] R. E. McKeighen and M. P. Buchin, "New techniques for dynamically variable electronic delays for real time ultrasonic imaging," in *1977 Ultrasonics Symposium*, 1977, pp. 250–254, doi: 10.1109/ULTSYM.1977.196834.

[2] R. J. Mailloux, "Phased array theory and technology," *Proceedings of the IEEE*, vol. 70, no. 3, pp. 246–291, 1982, doi: 10.1109/PROC.1982.12285.

[3] H. T. Friis and C. B. Feldman, "A multiple unit steerable antenna for short-wave reception," *Proceedings of the Institute of Radio Engineers*, vol. 25, no. 7, pp. 841–917, 1937, doi: 10.1109/JRPROC.1937.228354.

[4] P. J. Kahrilas, "HAPDAR—An operational phased array radar," *Proceedings of the IEEE*, vol. 56, no. 11, pp. 1967–1975, 1968, doi: 10.1109/PROC.1968.6773.

[5] P. E. Green, R. A. Frosch, and C. F. Romney, "Principles of an experimental large aperture seismic array (LASA)," *Proceedings of the IEEE*, vol. 53, no. 12, pp. 1821–1833, 1965, doi: 10.1109/PROC.1965.4453.

[6] Ö. Yilmaz, O. Yilmaz, and S. M. Doherty, "Migration," in *Seismic data analysis: processing, inversion, and interpretation of seismic data*, SEG Books, 2001, pp. 463–653.

[7] S. H. Gray, J. Etgen, J. Dellinger, and D. Whitmore, "Seismic migration problems and solutions," *Geophysics*, vol. 66, no. 5, pp. 1622–1640, 2001, doi: 10.1190/1.1487107.

[8] S. D. Pye, S. R. Wild, and W. N. McDicken, "Adaptive time gain compensation for ultrasonic imaging," *Ultrasound in Medicine & Biology*, vol. 18, no. 2, pp. 205–212, 1992, doi: 10.1016/0301-5629(92)90131-S.

[9] A. Kyriakou, E. Neufeld, B. Werner, M. M. Paulides, G. Szekely, and N. Kuster, "A review of numerical and experimental compensation techniques for skull-induced phase aberrations in transcranial focused ultrasound," *International Journal of Hyperthermia*, vol. 30, no. 1, pp. 36–46, 2014, doi: 10.3109/02656736.2013.861519.

[10] S. J. Sanabria, E. Ozkan, M. Rominger, and O. Goksel, "Spatial domain reconstruction for imaging speed-of-sound with pulse-echo ultrasound: simulation and in vivo study," *Phys. Med. Biol.*, vol. 63, no. 21, p. 215015, 2018, doi: 10.1088/1361-6560/aae2fb.

[11] E. Boni, A. C. H. Yu, S. Freear, J. A. Jensen, and P. Tortoli, "Ultrasound open platforms for next-generation imaging technique development," *IEEE Transactions on Ultrasonics, Ferroelectrics, and Frequency Control*, vol. 65, no. 7, pp. 1078–1092, 2018, doi: 10.1109/TUFFC.2018.2844560.

[12] J. Faurie, M. Baudet, J. Poree, G. Cloutier, F. Tournoux, and D. Garcia, "Coupling myocardium and vortex dynamics in diverging-wave echocardiography," *IEEE Trans Ultrason Ferroelectr Freq Control*, vol. 66, no. 3, pp. 425–432, 2019, doi: 10.1109/TUFFC.2018.2842427.

[13] O. M. H. Rindal, A. Austeng, A. Fatemi, and A. Rodriguez-Molares, "The effect of dynamic range alterations in the estimation of contrast," *IEEE Transactions on Ultrasonics, Ferroelectrics, and Frequency Control*, vol. 66, no. 7, pp. 1198–1208, 2019, doi: 10.1109/TUFFC.2019.2911267.

[14] F. Destrempes and G. Cloutier, "A critical review and uniformized representation of statistical distributions modeling the ultrasound echo

envelope," *Ultrasound in Medicine & Biology*, vol. 36, no. 7, pp. 1037–1051, 2010, doi: 10.1016/j.ultrasmedbio.2010.04.001.

[15] S. Salles, H. Liebgott, O. Basset, C. Cachard, D. Vray, and R. Lavarello, "Experimental evaluation of spectral-based quantitative ultrasound imaging using plane wave compounding," *IEEE Transactions on Ultrasonics, Ferroelectrics, and Frequency Control*, vol. 61, no. 11, pp. 1824–1834, Nov. 2014, doi: 10.1109/TUFFC.2014.006543.

[16] Jian-Yu Lu, "Experimental study of high frame rate imaging with limited diffraction beams," *IEEE Transactions on Ultrasonics, Ferroelectrics, and Frequency Control*, vol. 45, no. 1, pp. 84–97, 1998, doi: 10.1109/58.646914.

[17] D. Garcia, L. Le Tarnec, S. Muth, E. Montagnon, J. Porée, and G. Cloutier, "Stolt's f-k migration for plane wave ultrasound imaging," *IEEE Transactions on Ultrasonics, Ferroelectrics, and Frequency Control*, vol. 60, no. 9, pp. 1853–1867, 2013, doi: 10.1109/TUFFC.2013.2771.

[18] D. Garcia, L. Kadem, D. Savéry, P. Pibarot, and L.-G. Durand, "Analytical modeling of the instantaneous maximal transvalvular pressure gradient in aortic stenosis," *Journal of Biomechanics*, vol. 39, no. 16, pp. 3036–3044, 2006, doi: 10.1016/j.jbiomech.2005.10.013.

[19] G. Montaldo, M. Tanter, J. Bercoff, N. Benech, and M. Fink, "Coherent plane-wave compounding for very high frame rate ultrasonography and transient elastography," *IEEE Transactions on Ultrasonics, Ferroelectrics, and Frequency Control*, vol. 56, no. 3, pp. 489–506, 2009, doi: 10.1109/TUFFC.2009.1067.

[20] J. Porée, D. Posada, A. Hodzic, F. Tournoux, G. Cloutier, and D. Garcia, "High-frame-rate echocardiography using coherent compounding with Doppler-based motion-compensation," *IEEE Transactions on Medical Imaging*, vol. 35, no. 7, pp. 1647–1657, 2016, doi: 10.1109/TMI.2016.2523346.

[21] S. Bae, P. Kim, and T. Song, "Ultrasonic sector imaging using plane wave synthetic focusing with a convex array transducer," *The Journal of the Acoustical Society of America*, vol. 144, no. 5, pp. 2627–2644, 2018, doi: 10.1121/1.5065391.

[22] S. Shahriari and D. Garcia, "Meshfree simulations of ultrasound vector flow imaging using smoothed particle hydrodynamics," *Phys Med Biol*, vol. 63, pp. 1–12, 2018, doi: 10.1088/1361-6560/aae3c3.

[23] C. P. Loizou and C. S. Pattichis, "An overview of despeckle-filtering techniques," in *Handbook of speckle filtering and tracking in cardiovascular ultrasound imaging and video*, IET Digital Library, 2018, pp. 95–109.

[24] W. Burger and M. J. Burge, "Interpolation," in *Principles of digital image processing: core algorithms*, London: Springer-Verlag, 2009, pp. 210–237.

[25] T. L. Szabo, *Diagnostic ultrasound imaging: inside out*. Academic Press, 2004.

[26] D. C. M. Horvat, J. S. Bird, and M. M. Goulding, "True time-delay bandpass beamforming," *IEEE Journal of Oceanic Engineering*, vol. 17, no. 2, pp. 185–192, 1992, doi: 10.1109/48.126975.

[27] A. R. Selfridge, G. S. Kino, and B. T. Khuri-Yakub, "A theory for the radiation pattern of a narrow-strip acoustic transducer," *Appl. Phys. Lett.*, vol. 37, no. 1, pp. 35–36, 1980, doi: 10.1063/1.91692.

[28] D. Napolitano *et al.*, "Sound speed correction in ultrasound imaging," *Ultrasonics*, vol. 44, pp. e43–e46, 2006, doi: 10.1016/j.ultras.2006.06.061.

[29] E. L. Madsen, J. A. Zagzebski, R. A. Banjavie, and R. E. Jutila, "Tissue mimicking materials for ultrasound phantoms," *Medical Physics*, vol. 5, no. 5, pp. 391–394, 1978, doi: 10.1118/1.594483.

[30] C. Yoon, Y. Lee, J. H. Chang, T. Song, and Y. Yoo, "In vitro estimation of mean sound speed based on minimum average phase variance in medical ultrasound imaging," *Ultrasonics*, vol. 51, no. 7, pp. 795–802, 2011, doi: 10.1016/j.ultras.2011.03.007.

[31] F. Vignon *et al.*, "Fast frame rate 2D cardiac deformation imaging based on RF data: what do we gain?," in *2017 IEEE International Ultrasonics Symposium (IUS)*, 2017, pp. 1–4, doi: 10.1109/ULTSYM.2017.8092648.

[32] G. Y. Hou *et al.*, "Sparse matrix beamforming and image reconstruction for 2-D HIFU monitoring using harmonic motion imaging for focused ultrasound (HMIFU) with in vitro validation," *IEEE Transactions on Medical Imaging*, vol. 33, no. 11, pp. 2107–2117, 2014, doi: 10.1109/TMI.2014.2332184.

[33] J. R. Gilbert, C. Moler, and R. Schreiber, "Sparse matrices in Matlab: design and implementation," *SIAM J. Matrix Anal. Appl.*, vol. 13, no. 1, pp. 333–356, 1992, doi: 10.1137/0613024.

[34] M. Maggioni and T. Berger-Wolf, "Optimization techniques for sparse matrix–vector multiplication on GPUs," *Journal of Parallel and Distributed Computing*, vol. 93–94, pp. 66–86, 2016, doi: 10.1016/j.jpdc.2016.03.011.

[35] H. Liebgott, A. Rodriguez-Molares, F. Cervenansky, J. A. Jensen, and O. Bernard, "Plane-wave imaging challenge in medical ultrasound," *IEEE International Ultrasonics Symposium (IUS)*, pp. 1–4, 2016, doi: 10.1109/ULTSYM.2016.7728908.

[36] C. Madiena, J. Faurie, J. Porée, and D. Garcia, "Color and vector flow imaging in parallel ultrasound with sub-Nyquist sampling," *IEEE Trans Ultrason Ferroelectr Freq Control*, vol. 65, no. 5, pp. 795–802, 2018.

[37] B. Y. S. Yiu and A. C. H. Yu, "Least-squares multi-angle Doppler estimators for plane-wave vector flow imaging," *IEEE Transactions on Ultrasonics, Ferroelectrics, and Frequency Control*, vol. 63, no. 11, pp. 1733–1744, 2016, doi: 10.1109/TUFFC.2016.2582514.

[38] J. Yu, L. Lavery, and K. Kim, "Super-resolution ultrasound imaging method for microvasculature in vivo with a high temporal accuracy," *Scientific Reports*, vol. 8, no. 1, pp. 1–11, 2018, doi: 10.1038/s41598-018-32235-2.

[39] M. Imbault *et al.*, "Robust sound speed estimation for ultrasound-based hepatic steatosis assessment," *Phys Med Biol*, vol. 62, no. 9, pp. 3582–3598, 2017, doi: 10.1088/1361-6560/aa6226.

[40] R. E. Zubajlo *et al.*, "Experimental validation of longitudinal speed of sound estimates in the diagnosis of hepatic steatosis (part II)," *Ultrasound in Medicine & Biology*, vol. 44, no. 12, pp. 2749–2758, 2018, doi: 10.1016/j.ultrasmedbio.2018.07.020.

[41] R. Mallart and M. Fink, "Adaptive focusing in scattering media through sound-speed inhomogeneities: The van Cittert Zernike approach and focusing criterion," *The Journal of the Acoustical Society of America*, vol. 96, no. 6, pp. 3721–3732, 1994, doi: 10.1121/1.410562.

[42] B. Y. S. Yiu, I. K. H. Tsang, and A. C. H. Yu, "GPU-based beamformer: fast realization of plane wave compounding and synthetic aperture imaging," *IEEE Transactions on Ultrasonics, Ferroelectrics, and Frequency Control*, vol. 58, no. 8, pp. 1698–1705, 2011, doi: 10.1109/TUFFC.2011.1999.

[43] J. F. Synnevag, A. Austeng, and S. Holm, "Adaptive beamforming applied to medical ultrasound imaging," *IEEE Transactions on Ultrasonics, Ferroelectrics, and Frequency Control*, vol. 54, no. 8, pp. 1606–1613, 2007, doi: 10.1109/TUFFC.2007.431.

[44] B. M. Asl and A. Mahloojifar, "Eigenspace-based minimum variance beamforming applied to medical ultrasound imaging," *IEEE Transactions on Ultrasonics, Ferroelectrics, and Frequency Control*, vol. 57, no. 11, pp. 2381–2390, 2010, doi: 10.1109/TUFFC.2010.1706.

[45] M. Polichetti, F. Varray, J.-C. Béra, C. Cachard, and B. Nicolas, "A nonlinear beamformer based on p-th root compression—application to plane wave ultrasound imaging," *Applied Sciences*, vol. 8, no. 4, p. 599, 2018, doi: 10.3390/app8040599.

[46] J. Camacho, M. Parrilla, and C. Fritsch, "Phase coherence imaging," *IEEE Transactions on Ultrasonics, Ferroelectrics, and Frequency Control*, vol. 56, no. 5, pp. 958–974, 2009, doi: 10.1109/TUFFC.2009.1128.

[47] M. Polichetti, V. Perrot, H. Liebgott, B. Nicolas, and F. Varray, "Influence of beamforming methods on velocity estimation: in vitro experiments," in *2018 IEEE International Ultrasonics Symposium (IUS)*, 2018, pp. 1–4, doi: 10.1109/ULTSYM.2018.8580186.

[48] A. C. Luchies and B. C. Byram, "Deep neural networks for ultrasound beamforming," *IEEE Transactions on Medical Imaging*, vol. 37, no. 9, pp. 2010–2021, 2018, doi: 10.1109/TMI.2018.2809641.

[49] M. Gasse, F. Millioz, E. Roux, D. Garcia, H. Liebgott, and D. Friboulet, "High-quality plane wave compounding using convolutional neural networks," *IEEE Transactions on Ultrasonics, Ferroelectrics, and Frequency Control*, vol. 64, no. 10, pp. 1637–1639, 2017, doi: 10.1109/TUFFC.2017.2736890.

[50] J. Porée, M. Baudet, F. Tournoux, G. Cloutier, and D. Garcia, "A dual tissue-Doppler optical-flow method for speckle tracking echocardiography at high frame rate," *IEEE Trans Med Imaging*, vol. 37, no. 9, pp. 2022–2032, 2018, doi: 10.1109/TMI.2018.2811483.